\documentstyle[aps,prl,twocolumn,epsf]{revtex} 
\newcommand{\p}{\mbox{$^{\prime}$}} 
\newcommand{\znco}{\mbox{$\rm ZnCr_2O_4$}} 
\newcommand{\cpp}{\mbox{$\chi^{\prime\prime}$}} 
\begin{document} 
\input{psfig.sty} 
\draft 
\twocolumn[\hsize\textwidth\columnwidth\hsize\csname 
@twocolumnfalse\endcsname

\title{\bf  Local spin resonance and spin-Peierls-like phase transition\\ in a geometrically frustrated antiferromagnet} 
\author{S.-H.  Lee$^{1,2}$, C. Broholm$^{3,2}$, 
T.H. Kim$^{4}$\cite{bitter}, W. Ratcliff II$^4$ and S-W. Cheong$^{4,5}$} 
\address{$^{1}$ 
University of Maryland, College Park, Maryland 20742} 
\address{$^{2}$ 
NIST Center for Neutron Research, National Institute of Standards
and Technology, Gaithersburg, Maryland
20899} 
\address{$^{3}$ 
Department of Physics and Astronomy, The Johns Hopkins University,  
Baltimore, Maryland 21218} 
\address{$^4$Department of Physics and Astronomy, Rutgers University, Piscataway, New Jersey 08854} 
\address{$^5$Bell Laboratories, Lucent Technologies, Murray Hill, 
New Jersey 07974}  
\maketitle 
\begin{abstract} 
 
Inelastic magnetic neutron scattering reveals a localized spin
resonance at 4.5 meV in the ordered phase of the geometrically
frustrated cubic antiferromagnet $\rm ZnCr_2O_4$. The resonance
develops abruptly from quantum critical fluctuations upon cooling
through a first order transition to a co-planar antiferromagnet at
$T_c=12.5(5)$ K. We argue that this transition is a three dimensional
analogue of the spin-Peierls transition.

\end{abstract} 
 
 
]

\newpage

It appears that antiferromagnetically interacting Heisenberg spins on
the vertices of a lattice of corner-sharing tetrahedra cannot
order\cite{anderson,reimers,moes98,cana98}.  Characterized by weak
connectivity and frustrated interactions\cite{ramirev}, this so-called
pyrochlore antiferromagnet has no phase transition for spin
$S=\infty$\cite{reimers,moes98} and forms a quantum spin liquid with
low lying singlet excitations for $S=1/2$\cite{cana98}. Recent
experimental\cite{raju,jason} and theoretical\cite{bramwell,palmer}
work however indicates that small perturbations away from the ideal
model can induce low temperature phase transitions. In the quantum
critical spin-1/2 antiferromagnetic chain sensitivity to small symmetry
breaking perturbations leads to the  spin-Peierls transition: A lattice
distortion that lowers the energy of the spin system by inducing a gap
in the magnetic excitation spectrum.  In this letter we report a
similar phenomenon for spins on a lattice of corner-sharing
tetrahedra.

We examined the spinel antiferromagnet (AFM) \znco\ 
using neutron scattering.  
Cubic and paramagnetic at high temperatures, the material 
undergoes a first order phase transition at
$T=12.5$ K into a tetragonal phase with N\'{e}el order.  Our
measurements and analysis indicate that lattice energy is expended at
the phase transition to break the symmetry of frustrated magnetic
interactions and thereby select an ordered magnetic phase from a
manifold of degenerate states. A local spin resonance
develops abruptly below the transition. This resonance is a
manifestation of $Q$-space degeneracy in the ordered phase of a
frustrated magnet. 
Just as the development of a singlet-triplet gap drives the spin Peierls
transition, the fact that ordering in a frustrated magnet can push low
energy spectral weight into a finite energy resonance plays an important
role at the transition in \znco . 

Cr$^{3+}$ is the source of magnetism in \znco . Approximately
octahedrally coordinated, the unfilled 3d$^3$ shells of these atoms
form isotropic $S=3/2$ degrees of freedom on a lattice of corner-sharing
tetrahedra. Assuming that only nearest neighbor interactions are
appreciable yields an estimate of $J=(3k_B\Theta_{CW}/zS(S+1))=-4.5$
meV for the exchange constant. Here $\Theta_{CW}\approx-390$
K is the Curie-Weiss temperature derived from high temperature
susceptibility data\cite{ramirev} and $z=6$ is the nearest neighbor
coordination number.  Experiments on Cr$^{3+}$ pairs in $\rm
ZnGa_2O_4$\cite{gorkom,henning} give values for $J$ ranging from $-2.8$
meV to $-4.0$ meV and also provide evidence for biquadratic exchange
${\cal H}_{2}=j({\bf S}_1\cdot{\bf S}_2)^2$ with $j=-0.21(4)$ meV.

A  25 g powder sample was prepared by solid state reaction between 
stoichiometric amounts of Cr$_2$O$_3$ and ZnO in air.  Rietveld 
analysis of neutron powder diffraction data from the NIST BT1 
diffractometer shows that \znco\ in the spinel 
structure (space-group $Fd\bar{3}m$, $a=8.31273$\AA\ for $T=15$ K) 
is the majority phase with a minority phase of 
1\% f. u. un-reacted $\rm Cr_2O_3$. Elastic and inelastic neutron scattering 
measurements were performed at NIST on the cold neutron triple-axis 
spectrometer SPINS. A vertically focusing Pyrolytic Graphite (002) monochromator (PG(002)) 
extracted a monochromatic beam with energy 2.5 meV$<E_i<$ 
14 meV from a $^{58}$Ni coated cold neutron guide.  The detection 
system consisted of a 20 cm long polycrystalline BeO filter cooled to 
77 K followed by a 23 cm$\times$ 15 cm flat PG(002) analyzer 92 cm from 
the sample, then a $80\p$ radial collimator, and an area sensitive 
detector. The energy range detected was 2.6 meV$<E_f<$ 3.7 meV with 
Full Width at Half Maximum (FWHM) energy resolution 0.1 meV$<\Delta E<0.15$ 
meV and angular resolution $\Delta 2\theta\approx 50\p$. 
The absolute efficiency of the instrument was measured 
using incoherent elastic scattering from vanadium and nuclear
Bragg peaks from \znco. 
The corresponding normalization factor was applied to 
background subtracted data to obtain measurements of the normalized 
magnetic neutron scattering intensity\cite{lovesey} 
\begin{eqnarray} 
\tilde{I}(Q,\omega )=\int\frac{d\Omega_{\hat{\bf Q}}}{4\pi} 
|\frac{g}{2}F(Q)|^2 \sum_{\alpha\beta}(\delta_{\alpha\beta}-\hat{Q}_\alpha 
\hat{Q}_\beta ) {\cal S}^{\alpha\beta}({\bf Q},\omega )\nonumber . 
\end{eqnarray} 
Here $F(Q)$ is the magnetic form factor for Cr$^{3+}$\cite{xraytables} 
and

%
\vspace{0.15in} 
\noindent 
\parbox[b]{3.4in}{ 
\psfig{file=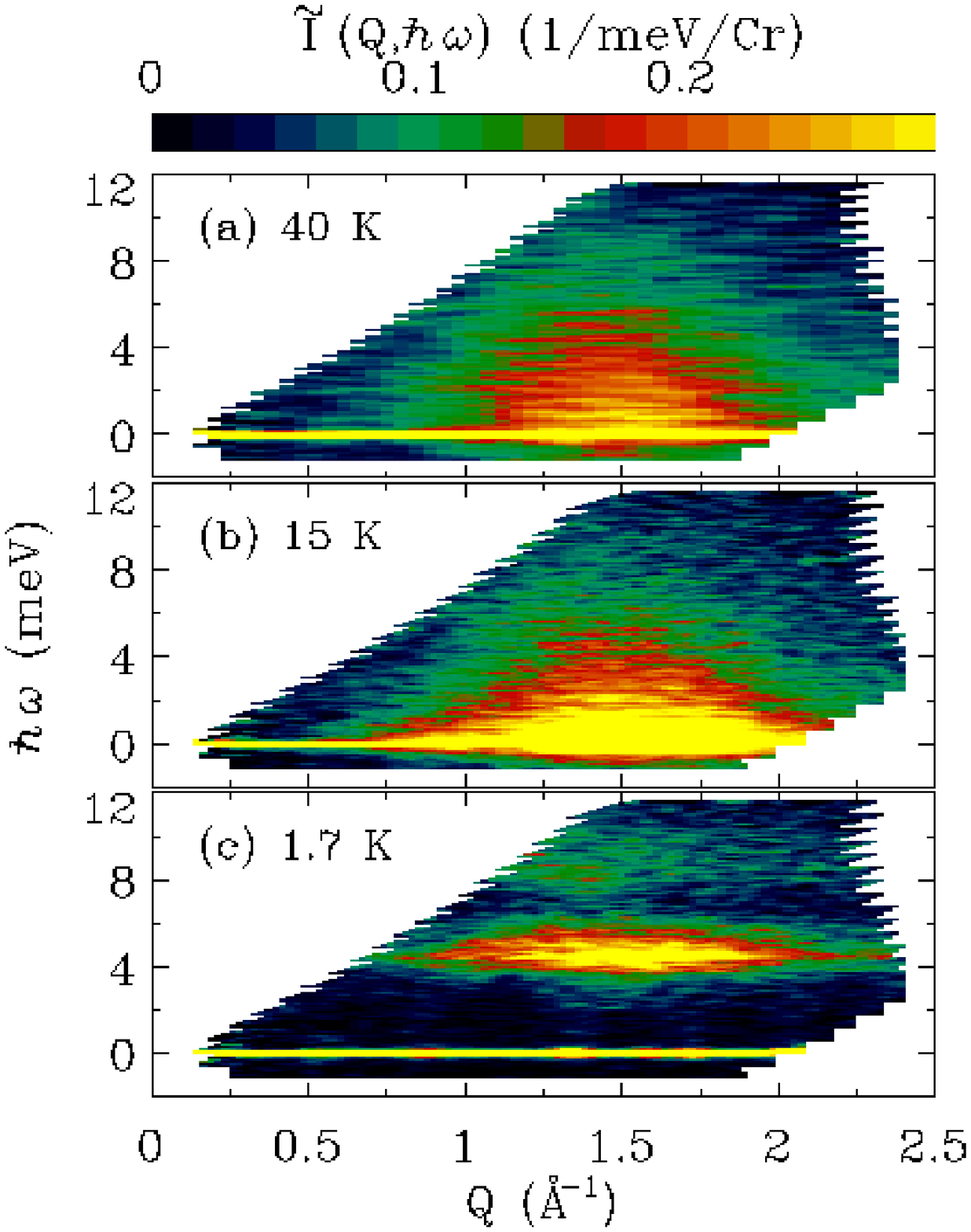,width=2.95in} 
{Fig.~1. \small 
Color maps of the magnetic neutron scattering intensity versus 
wave vector and energy transfer at three 
temperatures spanning the phase transition at $T_c=12.5(5)$~K.}} 
\vspace{0.05in}
%

\noindent
${\cal S}^{\alpha\beta}({\bf Q},\omega )$ is the scattering function\cite{lovesey}. 

Fig. 1 provides an overview of our data in the form of 
color images of $\tilde{I}(Q,\omega)$ at three temperatures.  In the 
paramagnetic phase,  Figs.~1 (a) and (b) 
provide evidence for quantum critical fluctuations of small 
AFM clusters most likely antiferromagnetically correlated 
tetrahedra. The data closely resemble those obtained in similar 
experiments on other frustrated 
AFM's\cite{jason,scgoneu}. For $T<T_c$ however, the low energy spectral weight concentrates into a 
sharp constant-energy mode centered at $\hbar\omega=4.5$ meV$\approx |J|>>k_BT_c$. 
Above the resonance there is an additional band 
of intensity centered at $\hbar\omega\approx 9$ meV (Fig. 2 (a)).   
Careful inspection of data below the resonance also  
reveals weak dispersing streaks emanating from AFM Bragg points.    
 
Focusing first on the ordered phase, Fig.~2 (a) shows 
the $Q-$integrated magnetic scattering cross section at $T=1.7$ K. 
With a resolution corrected FWHM of only 0.8 meV 
the $\hbar\omega=4.5$ meV peak is evidence for a near dispersionless excitation.
 The spectral weight is $\hbar\int^{6meV}_{3.5meV} \sum_{\alpha\alpha}{\cal 
S}^{\alpha\alpha}(\omega) d\omega = 0.59(1) ~/Cr$ which is 22\% of the 
total fluctuating moment ($(S(S+1)-|\langle{\bf S}\rangle|^2)=2.65(5)/Cr$) 
and corresponds to 16\% 
of the total magnetic scattering cross section. 
Fig.~2 (b) shows the Q-dependence 
of the energy integrated intensity of this mode. 
The data exhibit a broad peak centered at $Q_0=1.5$~\AA$^{-1}$ with a 
Half Width at Half Maximum $\kappa =$ 0.48(5)~\AA$^{-1}= 0.64(6) a^*$ 
indicating that the excited 
%
%
\vspace{0.15in} 
\noindent 
\parbox[b]{3.4in}{ 
\psfig{file=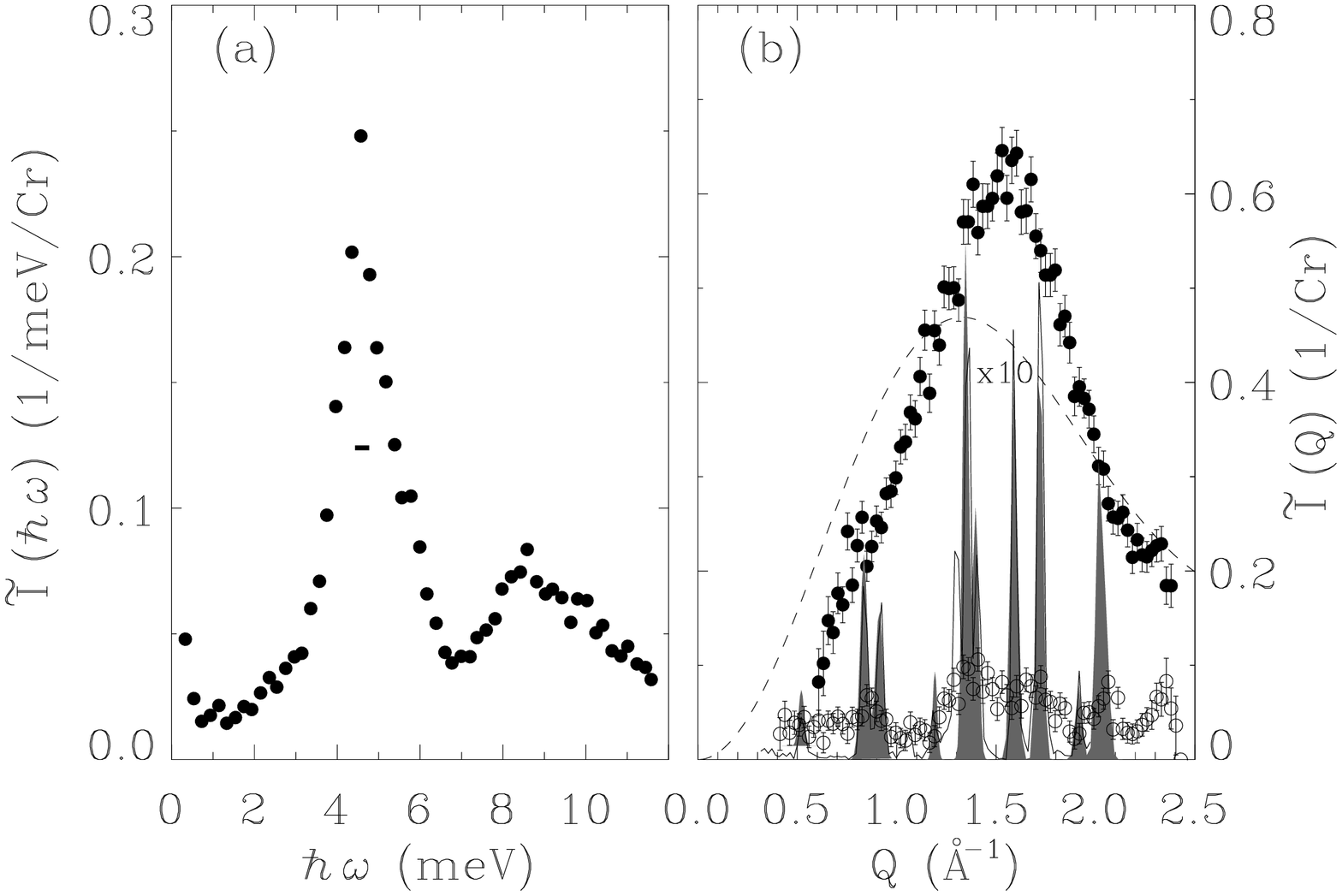,angle=90,width=3.4in} 
\vspace{0.05in} 
{Fig.~2. \small 
Integrated magnetic scattering intensities at $T=1.5$~K derived 
from Fig. 1 (c). 
(a) $\omega$-dependence of the $Q$-integrated 
intensity:  $\tilde{I}(\omega )=\int Q^2dQ \tilde{I}(Q,\omega )/\int 
Q^2 dQ$.  The horizontal bar shows the instrumental energy resolution. 
(b) Closed symbols show the Q-dependence of the $\omega$-integrated 
resonance intensity: 
$\tilde{I}(Q)=\hbar \int_{3.5~meV}^{6.0~meV}\tilde{I}(Q,\omega)d\omega$.  
Open circles show data integrated from 1 meV to 
3 meV.  The solid line shows the elastic scattering cross section 
scaled by a factor 10. Shaded peaks are magnetic Bragg peaks. 
}} 
%

\noindent 
state involves  
AFM correlated nearest neighbor spins. For comparison the dashed line shows 
the powder-averaged magnetic neutron scattering intensity 
for an isolated spin dimer at the nearest 
neighbor separation $r_0=2.939$~\AA \cite{furr79}.  
The spin pair model produces a broader peak than 
the experiment indicating that the resonating spin cluster 
in \znco\ is more complex.  
Fig.~2 (b) also shows the wave vector dependence of inelastic 
scattering below the resonance integrated over energy from 1 meV to 3 meV 
(open symbols). Weak, non-resolution-limited peaks are visible and their locations 
coincide  with AFM Bragg peaks (shaded). Excitations below the resonance are also apparent in the wave 
vector integrated data of Fig. 2 (a) where the intensity increases in 
proportion to energy for $\hbar\omega <3.5$ meV (the low energy upturn is incompletely 
resolved elastic scattering). Both features are consistent with 
neutron scattering from spin waves in a three dimensional 
AFM powder with a spin gap $\Delta< 1.5$ meV. In particular the ratio of 
elastic to inelastic scattering is consistent with estimates based on  
spin wave theory in the long wavelength limit. From the width of the peaks we estimate a spin wave velocity  
$v=18(2)$ meV\AA . This number is much less than the spin wave velocity 
for a bi-partite simple cubic AFM with $J=-2.8$~meV, $v=(2/\sqrt{3})z|J|S a=239$ meV\AA\, but only slightly larger than the  
spin wave velocity for a cubic AFM
with the critical temperature of \znco : 
$v=2\sqrt{3}k_BT_c\ a/(S+1)=12.4$ meV\AA . 

Spin wave theory  provides a useful starting point 
for understanding the resonance. Geometrical 
\vspace{0.15in} 
\noindent 
\parbox[b]{3.4in}{ 
\psfig{file=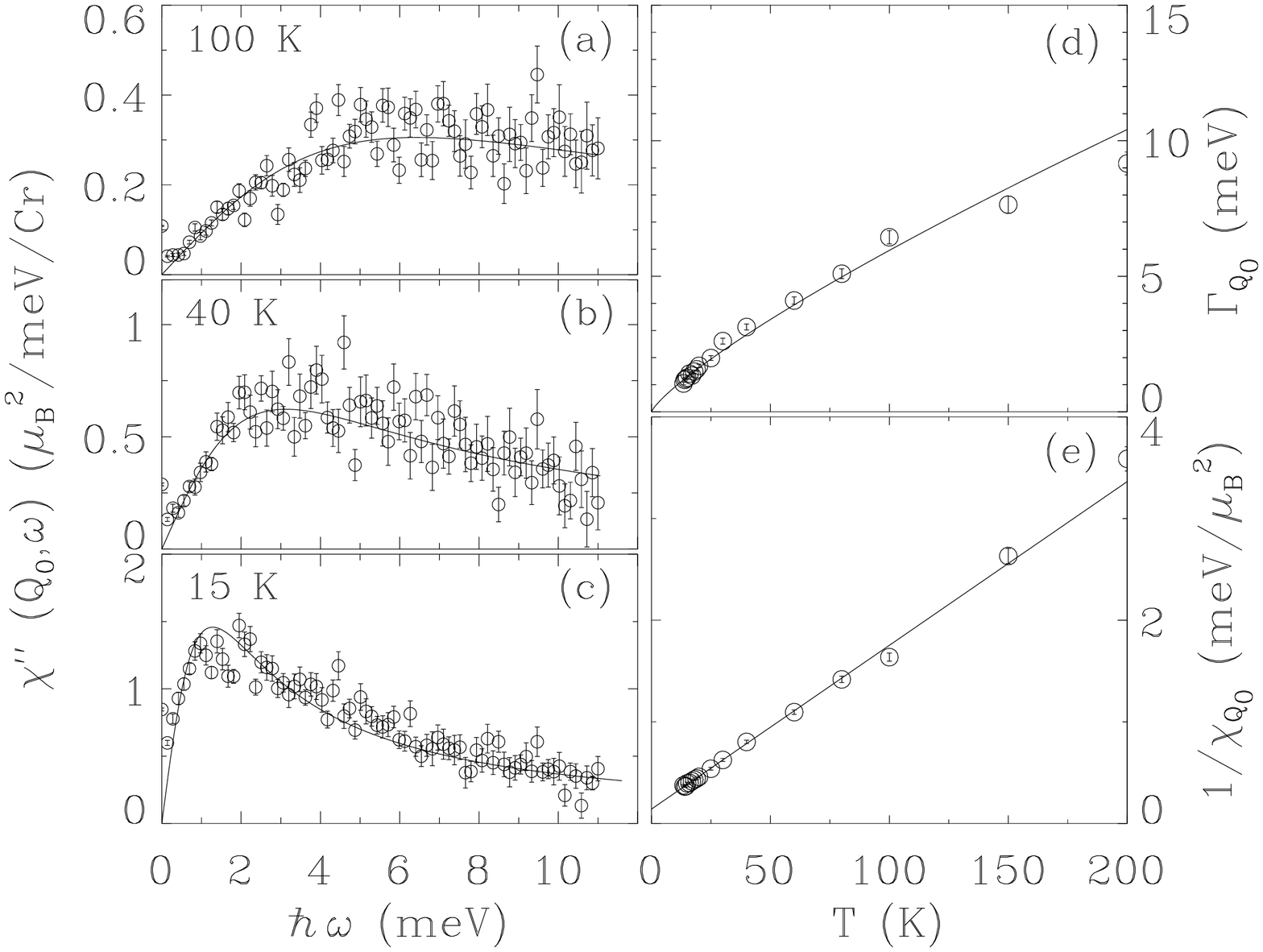,angle=90,width=3.4in} 
\vspace{0.05in} 
{Fig.~3. \small 
(a)-(c) $\chi(Q_0,\omega)$ at $Q_0=1.5$\AA$^{-1}$ derived from magnetic neutron 
scattering data via the fluctuation dissipation theorem. The solid lines  
are fits as described in the text. (d)-(e) Temperature dependence of the 
relaxation rate, $\Gamma_{Q_0} (T)$, and the inverse 
susceptibility, $\chi_{Q_0}^{-1}(T)$ derived from the fits. 
}} 

\noindent 
frustration 
leads to constant energy surfaces or volumes 
for spin wave dispersion 
relations in reciprocal space. Such $Q$-space ``degeneracy''   
in turn yields pronounced van-Hove singularities in wave vector 
averaged spectra.  Reimers {\em et al}\cite{reim91} showed 
that the pyrochlore AFM has two degenerate modes for 
any $\bf Q$ in the Brillouin zone. 
Wave vector independent excitations 
also exist for the kagom\'{e} AFM\cite{chubukov} 
and these have a real space interpretation 
in terms of the so-called weather-vane modes\cite{coleman}. 
A real-space interpretation has yet to be found for dispersionless 
excitations in the pyrochlore lattice.  
The broad peak in Fig. 2(b) indicates 
that they are highly localized in the ordered phase of \znco. 
 
Turning now to excitations in the paramagnetic phase, Fig.~3 
(a)-(c) show the imaginary part of the spin susceptibility, $\cpp (Q ,\omega 
)$  for several temperatures larger than $T_c$.  $\cpp (Q,\omega )$ was 
derived from inelastic neutron scattering data at $Q_0=1.5$ \AA$^{-1}$ via the fluctuation 
dissipation theorem:  $\chi\p\p(Q,\omega)=(g\mu_B)^2\pi 
(1-\exp(-\beta\omega )){\cal S}(Q,\omega)$. From the spectra we derived a 
temperature dependent spin relaxation rate, $\Gamma_Q $, and a static staggered 
susceptibility, $\chi_Q$,  by fitting to the 
following phenomenological response function: $ 
\chi\p\p(Q,\omega) = \chi_{Q} \Gamma_{Q}\omega/(\omega^2+\Gamma_{Q}^2)$. 
 
Figs. 3 (d) and (e) show the corresponding temperature dependent 
parameters. A power-law in $T$  describes the temperature dependence of 
$\Gamma_{Q_0} (T)={\cal C}\cdot k_BT(T/\theta)^{\alpha-1}$ for $T<150$ 
K while $\chi_{Q_0}(T)$ can be described by a Curie-Weiss law: 
$\chi_{Q_0}(T)=(\mu_{Q_0}^2/3k_B\theta)1/(1+T/\theta )$ in the entire 
temperature range. 
The best fit solid lines correspond to 
$\alpha=0.81(4)$,
 ${\cal C}=0.6(1)$, $\mu_{Q_0}=4.0(1)\mu_B$, and  
$\theta=8.8(4)$~K.  
Though we have plotted and analyzed data for 
$Q_0=1.5$~\AA$^{-1}$ we found similar results for $\Gamma (T)$ in 
analysis of wave-vector integrated data which probe the local spin susceptibility.  

\vspace{0.15in} 
\noindent 
\parbox[b]{3.4in}{ 
\psfig{file=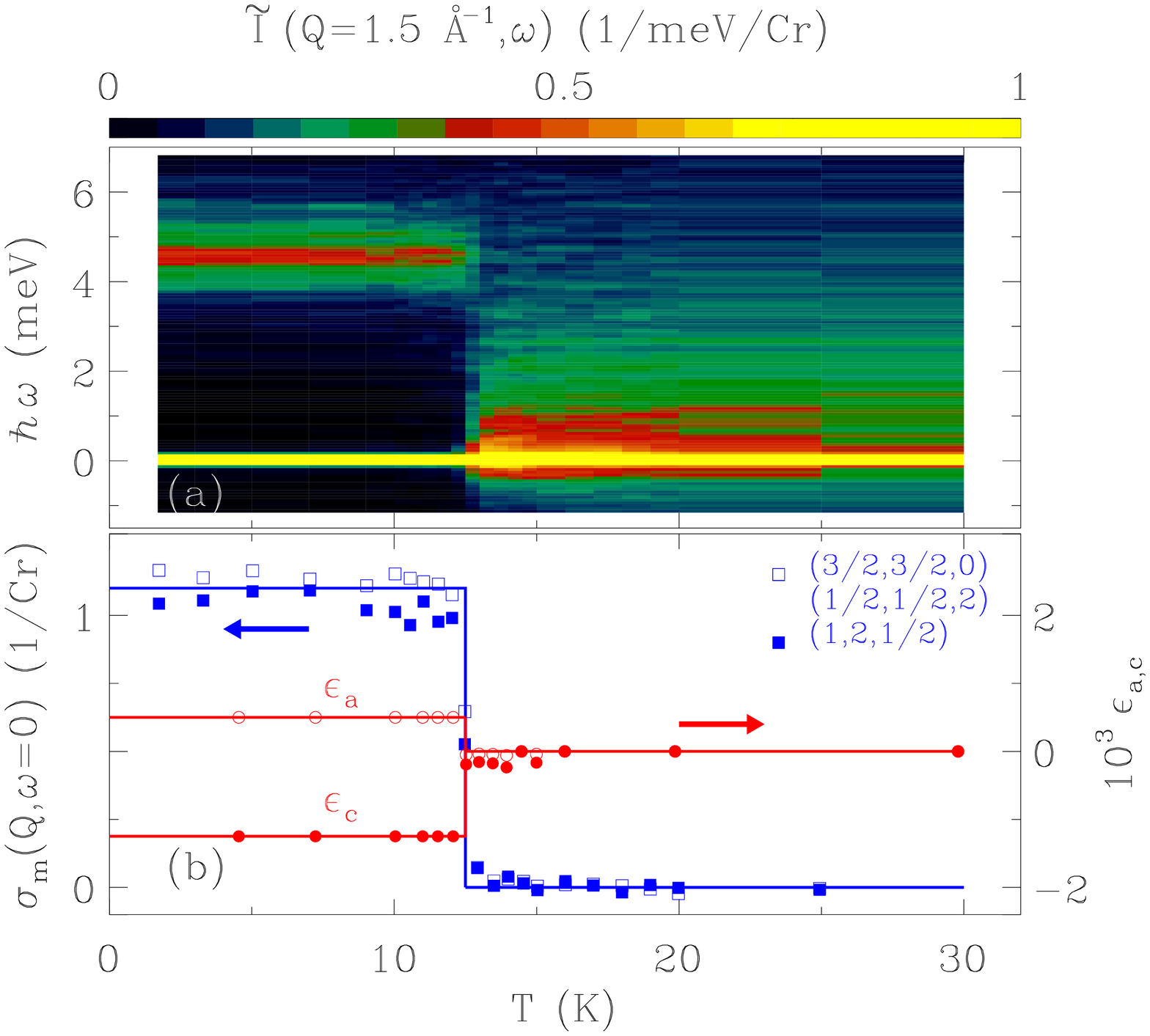,angle=90,width=3.1in} 
\vspace{0.05in} 
{Fig.~4. \small 
(a) Color image of inelastic neutron scattering
for $Q=1.5$~\AA$^{-1}$. 
(b)  $T-$dependence of magnetic Bragg scattering from a powder (blue squares),
$\sigma_m=\frac{v_m}{(2\pi)^3}
\int\tilde{I}(Q,\omega)4\pi Q^2dQd\hbar\omega$ where $v_m$ is the volume
per $Cr^{3+}$ ion,
and of lattice strain along $\bf a$ and $\bf c$ 
(red circles) measured by single crystal neutron diffraction. 
}} 
\vspace{0.1 in}

\noindent 
At a quantum critical point, $k_BT$ 
is the only low energy scale for local response functions.  If the
lorentzian form for $\chi\p\p(Q,\omega)$ describes the spectrum, then
$\alpha$ must equal one at the critical point. This was the exponent
found in Monte Carlo simulations of classical spins on a pyrochlore
lattice\cite{moes98}.  The deviation of $\alpha$  from unity in
\znco\ is perhaps not surprising given that the material does exhibit a
magnetic phase transition.  However, the fact that $\Gamma_Q$ tends to
zero as $T\rightarrow 0$ rather than at $T_c$ indicates that the system
may actually be approaching a quantum disordered phase with a gap
$k_B\theta =0.75$ meV before being interrupted by a first order
transition to an unrelated competing phase.  This idea is consistent
with Fig. 4 which compares the temperature dependent lattice strain and
magnetic Bragg peak intensity (frame (b)) to the inelastic neutron
scattering spectrum at $Q_0=1.5$~\AA$^{-1}$ (frame (a)).  Magnetic Bragg
peaks, a tetragonal lattice distortion, and the spin resonance all
appear abruptly, and without conventional critical fluctuations at
$T_c$.
 
Theoretical work has shown that magnetic order can not develop in an
isotropic spin pyrochlore AFM\cite{moes98,cana98}. There are many
possible deviations from the perfect model that could cause \znco\ to
order nonetheless. These include further neighbor interactions and spin
space anisotropy\cite{raju,bramwell,palmer}.  
Because the transition in \znco\ is
of the first order and involves a lattice distortion, we suggest that
finite lattice rigidity is an important factor at the phase transition
in this material.  Consider the effect of tetragonal strain on
magnetism in \znco .  It is well known that the exchange interaction
between Cr$^{3+}$ ions whose oxygen coordination octahedra share an
edge is strongly dependent on the Cr-Cr spacing, $r$\cite{good}.
Analysis of a series of chromium oxides indicates that $dJ/dr\approx
40$ meV/\AA \cite{motida}.  This implies that tetragonal strain
$\epsilon_a>0$ and $\epsilon_c<0$ yields weaker AFM interactions in the
basal plane, $\Delta J_{\perp}=r_0 \epsilon_a dJ/dr=0.06$ meV, and
stronger AFM interactions between all other spin pairs $\Delta
J_{\parallel}=r_0 (\epsilon_a+\epsilon_c)/2 dJ/dr=-0.04$ meV.  This
asymmetry reduces the mean field energy of the ordered phase in
\znco\ by $\Delta\langle \overline{\cal H}_s \rangle=(5\Delta
J_{\parallel}-\Delta J_{\perp} )/2\approx -0.07$ meV/Cr relative to the
mean field ground state energy in the cubic phase\cite{oles70}. The
result should be N\'{e}el order in tetragonal \znco\ below an ordering
temperature that we denote $T_{Nt}$.  Because order appears abruptly
and simultaneously with the tetragonal strain we infer that
$T_{Nt}>T_c$.  This may be possible despite the modest value of
$\Delta\langle \overline{\cal H}_s \rangle $ because of the strong
local constraints present when $|T/\Theta_{CW}|<<1$. In addition there
could be other hitherto undetected lattice modifications at $T_c$ that
also favor N\'{e}el order (see below).

The magnitude of the lattice distortion is controlled 
by the need to balance the increase in lattice energy
and the decrease in entropy
against the decrease in the energy of the spin system.  
Equating the free energy 
$F=\langle{\cal H}_l+{\cal H}_s\rangle-TS$ of the competing  
phases at $T_c$ implies that  
\begin{equation} 
\Delta\langle{\cal H}_s\rangle+\Delta\langle{\cal H}_l\rangle-T_c\Delta S=0 
\end{equation} 
 when cooling through $T_c$. 
We can derive $\Delta\langle{\cal H}_s\rangle$  
from Fig. 1 (b) and (c) using the first moment sum rule:\cite{hohen}
\begin{equation} 
\Delta\langle{\cal H}_s\rangle = 
-\frac{3}{2}\frac{\hbar^2\int_0^{\infty}\omega (1-e^{-\beta\hbar\omega}) 
\Delta{\cal S}(Q,\omega ) d\omega}{1-\sin Qr_0/Qr_0} 
\end{equation} 
Limiting the integral to 0.2 meV to 12 meV and averaging data for  
$1.3 ~{\rm \AA}^{-1} < Q < 2 ~{\rm \AA}^{-1}$ yields a 
value of $\Delta\langle{\cal H}_s\rangle=-0.40(7)$ meV/Cr.
From specific heat measurements\cite{ramirev} we find that 
$\Delta S=-0.107~R\ln 4$/Cr corresponding to $-T_c\Delta S=0.16$ meV/Cr.
The difference between these numbers yields an estimate
for the increase in lattice energy 
$\Delta\langle{\cal H}_l\rangle=0.24(7)$ meV/Cr. 
The energy associated with simple 
tetragonal strain\cite{kino} only accounts for 
$v_0 (c_{11}(\epsilon_c^2+2\epsilon_a^2)/2+c_{12}
(\epsilon_a^2+2\epsilon_a\epsilon_c))=0.026$ meV/Cr
so there are likely additional
modifications to the structure of \znco\ below $T_c$ that help to 
stabilize N\'{e}el order.
 
There are interesting analogies between the phase transition in
\znco\ and the spin-Peierls transition.  In both cases the high $T$
phase is near quantum critical and can lower its energy through a
lattice distortion. In both cases the transition occurs from a strongly
correlated paramagnet: $T_c\ll\Theta_{CW}$.  And in both cases low
energy spectral weight is moved into a finite energy peak. There are
also important differences that render the transition in \znco\ a
distinct new phenomenon in magnetism.  The lattice distortion in
\znco\ drives the spin system into an ordered phase not a quantum
disordered phase. The transition in \znco\ is of the first order while
the SP transition is of the second order. And the change in entropy at
$T_c$ plays an important role in \znco , not in a SP transition.  The
central idea that finite lattice rigidity can drive a spin system away
from quantum criticality  however does carry over,
and might be relevant for other frustrated magnets when symmetry
breaking terms in the spin Hamiltonian fail to induce magnetic order.

We thank A. P. Ramirez for helpful discussions and for access to
unpublished specific heat data.  The NSF supported work at SPINS
through DMR-9423101, work at JHU through DMR-9453362 and work at
Rutgers through DMR-9802513.

\end{document}